%% file: main.tex
\documentclass[superscriptaddress,twocolumn,aps,pra]{revtex4-2}

\usepackage{amsmath,amssymb,color,graphicx,dcolumn,bm,verbatim,physics,ulem}
\usepackage{lipsum}
\usepackage{xcolor}
\usepackage[utf8]{inputenc}
\usepackage[english]{babel}
\usepackage{natbib}
\usepackage{enumitem}
\usepackage{qcircuit}
\usepackage{verbatim}
\usepackage[hidelinks]{hyperref}

\def\ket#1{\left|#1\right>}
\def\<{\langle}
\def\>{\rangle}

\definecolor{deepcarrotorange}{rgb}{0.91, 0.41, 0.17}
\definecolor{cadmiumgreen}{rgb}{0.0, 0.62, 0.24}

\definecolor{darkpastelgreen}{rgb}{0.01, 0.75, 0.24}

\def\injfiddone{$98.9_{-0.1}^{+0.1}\%$}

\def\injfiddthreeraw{$94.1_{-0.1}^{+0.1}\%$}

\def\injfiddthreeps{$98.3_{-0.1}^{+0.1}\%$}

\def\injfiddthree{$95.1_{-0.1}^{+0.1}\%$}

\def\distfiddthree{$99.4_{-0.4}^{+0.3}\%$}

\def\injfiddfive{$92.5_{-0.2}^{+0.1}\%$}

\def\distfiddfive{$98.6_{-1.3}^{+0.9}\%$}

\def\gainfiddfive{$6.1_{-1.4}^{+0.9}\%$}

\begin{document}

\title{Experimental 
demonstration of logical magic state distillation}

\include{authors}

\begin{abstract}
Realizing universal fault-tolerant quantum computation is a key goal in quantum information science~\cite{shor1996fault,aharonov1999fault,gottesman2010introduction,campbell2017roads}.
By encoding quantum information into logical qubits utilizing quantum error correcting codes, physical errors can be detected and corrected, enabling substantial reduction in logical error rates~\cite{acharya2024quantum,acharya2023suppressing,ryan-anderson2022implementing,bluvstein2024logical,putterman2024hardware,sivak2023real,paetznick2024demonstration}.
However, the set of logical operations that can be easily implemented on such encoded qubits is often constrained~\cite{eastin2009restrictions,shor1996fault}, necessitating the use of special resource states known as `magic states'~\cite{bravyi2005universal} to implement universal, classically hard circuits ~\cite{gottesman1998heisenberg}. 
A key method to prepare high-fidelity magic states is to perform `distillation', creating them from multiple lower fidelity inputs~\cite{nielsen2010quantum,bravyi2005universal}.
Here we present the experimental realization of magic state distillation with logical qubits on a neutral-atom quantum computer. Our approach makes use of a dynamically reconfigurable architecture~\cite{bluvstein2022quantum,bluvstein2024logical} to encode and perform quantum operations on many logical qubits in parallel.
We demonstrate the distillation of magic states encoded in $d\,{=}\,3$ and $d\,{=}\,5$ color codes, observing improvements of the logical fidelity of the output magic states compared to the input logical magic states.
These experiments demonstrate a key building block of universal fault-tolerant quantum computation, and represent an important step towards large-scale logical quantum processors.
\end{abstract}

\maketitle

Quantum error correction (QEC) enables scalable quantum computation by exponentially suppressing logical error rates.
However, the set of logical operations that can be efficiently implemented on these encoded qubits is constrained, making it challenging to perform universal quantum processing~\cite{eastin2009restrictions}. 
For example, many QEC codes only support the realization of so-called Clifford gates~\cite{bravyi2012classification,fowler2012surface,bombin2006topological}.
Since Clifford gates can be efficiently simulated classically~\cite{gottesman1998heisenberg}, additional non-Clifford resources are required to achieve computational universality and quantum advantage.
To circumvent this obstacle, special quantum states, aptly named ``magic states", can be utilized to complete a universal set of logical operations via gate teleportation~\cite{nielsen2010quantum}. Due to their relatively high cost of preparation, these magic states are often considered the key resource for scalable processing~\cite{campbell2017roads}. 

\begin{figure*}
    \centering
    \includegraphics[width=1.0\linewidth]{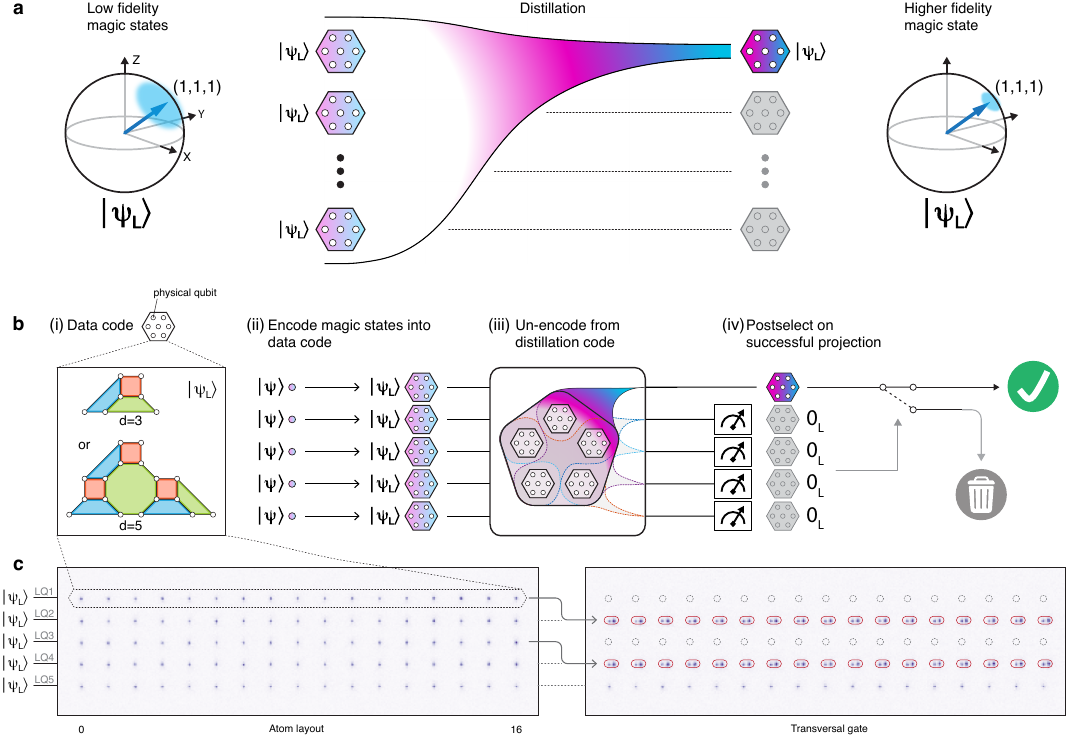}
    \caption{
    \textbf{Logical magic state distillation factory.}
    \textbf{a,} Schematic overview: Bloch sphere representation of magic state $\ket{\psi_L}$ (left) pointing in the (1,1,1) direction with shaded region indicating noise. Distillation~(center) takes multiple noisy logical inputs and produces a higher fidelity magic state~(right).
    \textbf{b,} 5-to-1 distillation procedure (left to right). Non-fault-tolerant encoding of physical magic states into five \textit{data code} logical qubits $\ket{\psi_L}$ protects 
    logical operations (i and ii). In particular, we encode into distance 3 and 5 color codes~(i).
    Encoded states (ii) are purified using a \textit{distillation code}. By running the un-encoding circuit of the distillation code (iii) and conditioning on distillation syndromes (iv), we have simultaneously projected into the code state of the distillation code and un-encoded it into the output magic state.
    Upon measuring the correct distillation syndromes, the output qubit has been ``distilled'' to a higher fidelity along the (1,1,1) direction.
    \textbf{c,} Averaged atom images from the $d\,{=}\,5$ distillation experiment, showing 85~physical qubits encoded into 5~logical qubits (LQ1 to LQ5) with 17~physical qubits each (left), shown here in SLM traps. Rows of logical qubits are coherently reconfigured for transversal CZ gates throughout the distillation circuit (right), shown here with LQ1 and LQ3 in AOD traps.
    }
    \label{fig:architecture}
\end{figure*}

High-fidelity magic states can be produced by refining multiple noisy copies through \textit{magic state distillation} (MSD)~\cite{bravyi2005universal}.
The noisy states, encoded in \textit{data} QEC codes, are concatenated into a \textit{distillation} code and purified through protected operations on the data codes (Fig.~\ref{fig:architecture}).
If the input fidelity exceeds the so-called distillation threshold, the fidelity of the output state is improved compared to the input.
An attractive feature of MSD is that the logical-level circuit is independent of the data QEC code.
Consequently, by increasing the data code distance and implementing the logical circuit with encoded operations, the logical fidelity of the output magic states can be improved, in principle, to any desired level through multiple rounds of distillation~\cite{litinski2019magic,gidney2019efficient,ogorman2017quantum}.

Important recent experiments have demonstrated MSD with physical qubits~\cite{souza2011experimental,brown2023advances}, but the direct physical encoding prevents suppression of logical gate errors.
Complementary experiments have also shown remarkable progress in error-suppressed encoding of magic states into logical qubits using flag protocols~\cite{postler2022demonstration,ye2023logical,gupta2024encoding}. 
However, without the protection provided by the data code, these approaches generally have higher complexity and low success probability when targeting very low logical error rates, although recent work has significantly improved their performance for low physical error rates~\cite{chamberland2019fault,gidney2024magic}.

We realize magic state distillation at the logical level on a neutral atom quantum computer.
Magic states are encoded using 2D color codes~\cite{bombin2006topological}, and subsequently a 5-to-1 logical magic state distillation is performed~\cite{bravyi2005universal}.
The factory outputs a single magic state and the remaining four qubits, which we call \textit{distillation syndromes}, determine successful distillation.
Central to our approach is the dynamic reconfigurability and high degree of parallel control of the neutral atom processor~\cite{bluvstein2022quantum,bluvstein2024logical}.
We realize gate and layout-efficient encoding circuits for arbitrary logical states in the $d\,{=}\,3$ ($d\,{=}\,5$) color codes, executing 10 (5) logical qubit encoding circuits in parallel.
MSD is carried out using transversal Clifford gates, efficiently implemented with parallel atom rearrangement across all code distances.
Correlated decoding~\cite{cain2024correlated,bluvstein2024logical} is applied to the distillation syndromes, and their stabilizer values are further leveraged as flags~\cite{chao2018fault,chamberland2019fault} to enhance output logical fidelity.
The operation of the magic state distillation factory is verified by 
distilling states with varying input fidelity, and confirming the error suppression scaling.
Conditioned on observing the correct logical outcome and suitable stabilizer patterns on the four distillation syndrome logical qubits, we obtain an enhancement of logical magic state fidelity from \injfiddthree{} to \distfiddthree{} for $d\,{=}\,3$ and from \injfiddfive{} to \distfiddfive{} for $d\,{=}\,5$, respectively.
This corresponds to a factor of $8_{-3}^{+8}$ infidelity suppression for $d\,{=}\,3$ and a factor of $6_{-3}^{+10}$ for $d\,{=}\,5$.

\begin{figure}
    \centering
    \includegraphics[width=1.0\linewidth]{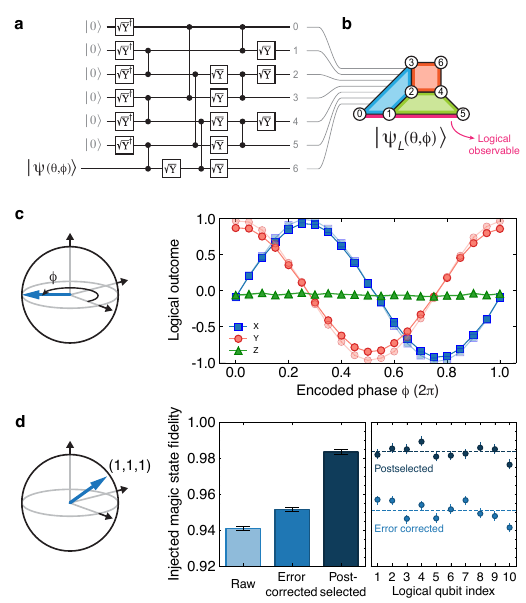}
    \caption{
    \textbf{Parallel logical encoding of arbitrary states.}
    \textbf{a,} Circuit for injecting an arbitrary state $\ket{\psi(\theta,\phi)}$ into the [[7,1,3]] color code.
    \textbf{b,} Schematic of $d\,{=}\,3$ color code stabilizers indicated by the three colored regions, with a logical operator highlighted. 
    \textbf{c,} Bloch sphere representation of the injected state with varying angle $\phi$ on the XY plane (left). Error-corrected logical outcomes for X, Y, Z measurement basis versus the injected phase. Faded markers indicate outcomes upon postselection on perfect stabilizers.
    \textbf{d,} Left, Bloch sphere representation of the (1,1,1) magic state. Center, Injected $d\,{=}\,3$ magic state fidelity corresponding to raw, error-corrected and postselected on perfect stabilizers, averaged across all 10 logical qubits.
    Right, Spatial distribution of injected magic state fidelities.}
    \label{fig:d3_colorcode}
\end{figure}

\section*{Arbitrary logical state encoding}

Our experiments utilize a newly built, Gemini-class quantum processor built and operated at QuEra~\cite{robinson2024benchmarking}. Inspired by earlier experiments from Harvard~\cite{bluvstein2024logical}, it involves control over a two-dimensional array of neutral atom qubits in a reconfigurable architecture.

We start by preparing magic states encoded in the data QEC code, with fidelity above the distillation threshold (for 5-to-1 distillation, the threshold fidelity for depolarizing errors is 83\%~\cite{bravyi2005universal}).
We choose the 2D color code as our data QEC code, as the full Clifford group can be implemented transversally with it~\cite{bombin2006topological}. The [[7,1,3]] color code is illustrated in Fig.~\ref{fig:d3_colorcode}b, where an X and Z stabilizer is associated with each colored region, and the logical operators lie along the edge.
Errors flip stabilizer values, so measuring stabilizers allows us to detect and correct physical errors in the circuit.
The parameters $[[n,k,d]]$ denote a QEC code with $n$ physical data qubits, $k$ logical qubits, and code distance $d$, which can correct $\lfloor\frac{d-1}{2}\rfloor$ errors or detect $d\,{-}\,1$ errors.
We use an arbitrary state encoding circuit that takes a physical qubit as input and encodes its state into a logical qubit (also known as state injection).
In particular, to encode into the $d\,{=}\,3$ 2D color code, we use the circuit in Fig.~\ref{fig:d3_colorcode}a, optimized for atom movement and number of entangling gate layers (Methods).

We verify the encoding circuit by injecting a state lying on the X$\mathrm{-}$Y plane and varying its angle (Fig.~\ref{fig:d3_colorcode}c).
This results in a rotation of the encoded logical information, which can be read out as an oscillation in logical measurements in the X or Y basis.
Logical measurements in the Pauli (X,Y,Z) basis are performed transversally, by measuring each physical qubit in the corresponding basis.
To interpret data qubit measurement results, we calculate the stabilizer and logical outcome parities from physical measurement results. 
If errors are detected, we can perform error correction on the logical result or alternatively discard the measurement (error detection).

With the ability to encode arbitrary states, we shift our focus to encoding magic states for further use in distillation.
In this work, we encode magic states that point in the (1,1,1) direction on the Bloch sphere, for use in the MSD procedure based on the [[5,1,3]] code~\cite{bravyi2005universal}. 
We prepare this state by initializing in $\ket{0}$ followed by a local single-qubit rotation of angle $\arccos{(1/\sqrt{3})}$ about the $(-1,1,0)$ axis on the physical qubit to be injected.
We perform logical quantum state tomography to estimate magic state fidelity (Methods).
We find that the encoded logical magic states have raw logical fidelity \injfiddthreeraw~(no error correction), error-corrected logical fidelity \injfiddthree, and error-detected logical fidelity \injfiddthreeps~(postselect on perfect stabilizers), Fig.~\ref{fig:d3_colorcode}d.
The error-detected state fidelity is close to the original physical magic state fidelity of \injfiddone{}, indicating that most of the added errors during the encoding process will also trigger syndromes.
To scalably use the resource states in a larger circuit, we cannot rely on postselecting on stabilizers that only become available when performing transversal measurements of the logical qubit later on.
Therefore, we focus on comparing the logical fidelity of magic states when only error correction (and no further postselection) is applied on the target magic state.

\begin{figure*}
    \centering
    \includegraphics[width=1.0\linewidth]{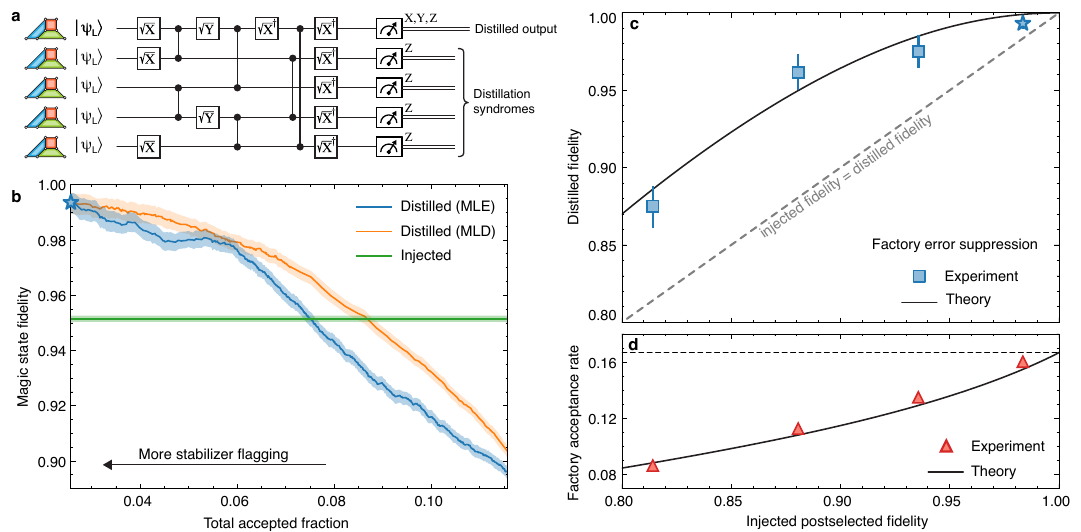}
    \caption{\textbf{5-to-1 magic state distillation.}
    \textbf{a,} Magic state distillation circuit based on the [[5,1,3]] code (distillation code). We measure distillation syndromes in the Z basis and perform tomography on the distilled output. The successful distillation syndrome for this circuit is 1011 (Methods).
    \textbf{b}, Fidelity of the output magic state for the $d\,{=}\,3$ distillation (blue line for the MLE decoder, orange line for the MLD decoder, see main text) as a function of the total accepted fraction, which includes both sliding scale postselection on distillation syndrome stabilizers, and the factory acceptance (1/6 in the noiseless case). With sufficient stabilizer flagging, the output fidelity exceeds that of the input error-corrected magic state fidelity (green). The shaded regions indicate 68\% confidence intervals, equivalent to $1\sigma$.
    \textbf{c}, We examine the distilled fidelity with full stabilizer postselection, after introducing coherent Z errors to the input magic states ($0.32\pi$, $0.24\pi$, $0.16\pi$ and $0$, left to right, blue points). 
     The results are in good agreement with the theoretical expectation (gray line). The stars in b and c indicate the same data point.
    \textbf{d}, Factory acceptance rate of distillation syndromes after perfect stabilizer postselection with the same coherent errors as panel \textbf{c}. Dashed line indicates the $1/6$ acceptance rate of the 5-to-1 magic state factory in the noiseless case.
    \label{fig:d3_distillation}}
\end{figure*}

\section*{5-to-1 Magic State Distillation}
The logical encoding circuit described above is not fault-tolerant, since physical errors on the injected physical qubit will lead to logical errors, resulting in a logical error rate that scales linearly with the physical error rate.
To further suppress the logical error rate, we make use of magic state distillation, which uses the properties of a \textit{distillation} QEC code and the fault-tolerant gates of the \textit{data} QEC code to improve the magic state quality (Fig.~\ref{fig:architecture}).

Our magic state factory is based on the [[5,1,3]] perfect code~\cite{laflamme1996perfect,bravyi2005universal}.
Schematically, the factory takes five noisy logical magic states as input and applies a unitary un-encoding circuit of the distillation [[5,1,3]] code, which we optimize to have only three layers of entangling gates (Fig.~\ref{fig:d3_distillation}a, Methods).
Measuring four of these logical qubits effectively measures the stabilizers of the distillation code, while the remaining logical qubit contains the output magic state.
By postselecting on the appropriate logical outcome of the four logical qubits, we achieve quadratic suppression of the logical error rate.
In the absence of errors, the factory acceptance rate of the 5-to-1 distillation factory is expected to be $1/6$ ~\cite{bravyi2005universal}.

To decode the logical measurement results, we use a most-likely error (MLE) correlated decoder based on mixed-integer programming~\cite{cain2024correlated}, with error weights obtained from separate characterization of our system (Methods).
For $d=3$, we also explore a maximum likelihood decoder (MLD) that simulates the majority logical outcome for a given stabilizer pattern and uses it to determine the correction (Methods).
These decoders can also be used to characterize the confidence of a given logical outcome assignment, allowing further sliding-scale postselection based on observed stabilizer patterns~\cite{bluvstein2024logical,smith2024mitigating,bombin2022fault} in analogy to flag protocols~\cite{chao2018quantum,chamberland2019fault,ryan-anderson2021realization,postler2022demonstration,gupta2024encoding}.
This postselection is commonly employed in theoretical analysis when preparing resource states, and in accordance with this, we use only the stabilizers of the four distillation syndrome logical qubits to perform decoding for postselection, since the output logical qubit is meant to be used for subsequent operations.

Experimental results of our logical magic state distillation factory are shown in Fig.~\ref{fig:d3_distillation}b.  Starting with the error-corrected input logical magic state with fidelity \injfiddthree{}, without any stabilizer postselection, the output magic state fidelity is worse than the injected state, due to the added physical errors during the distillation process.
However, we find that approximately 50\% stabilizer postselection is sufficient to improve the output magic state fidelity, and full postselection on perfect stabilizers of distillation syndrome qubits results in a fidelity of \distfiddthree{}.
Both decoders show similar performance, with the MLD decoder performing slightly better by accounting for the entropy of error configurations.

We further probe the physics of error suppression of the distillation code by artificially introducing coherent errors across the five input logical qubits (Fig.~\ref{fig:d3_distillation}c,d), which we achieve by applying a Z rotation on the physical qubits prior to state injection.
After encoding, this results in a magic state rotated around the X axis, which we use as input to the factory.
After full stabilizer postselection on the four distillation syndrome qubits,
we compare the error-corrected output fidelity against the postselected input fidelity, in order to highlight the distillation behavior on the logical information.
We observe distillation gain for all added rotation angles.

As the added rotation angle error increases, we observe that the output state infidelity is consistent with quadratic suppression of the added error.
We also find that the factory acceptance rate decreases with added errors, with an initial decrease that scales linearly with the added error.
This can be understood from the fact that a single input logical error will lead to an outcome different from the correct distillation syndrome, reducing the factory acceptance rate without contributing to the distilled fidelity.
Two input logical errors are needed to affect the distilled fidelity, giving rise to quadratic error suppression~\cite{bravyi2005universal}.

\begin{figure}
    \centering
    \includegraphics[width=\linewidth]{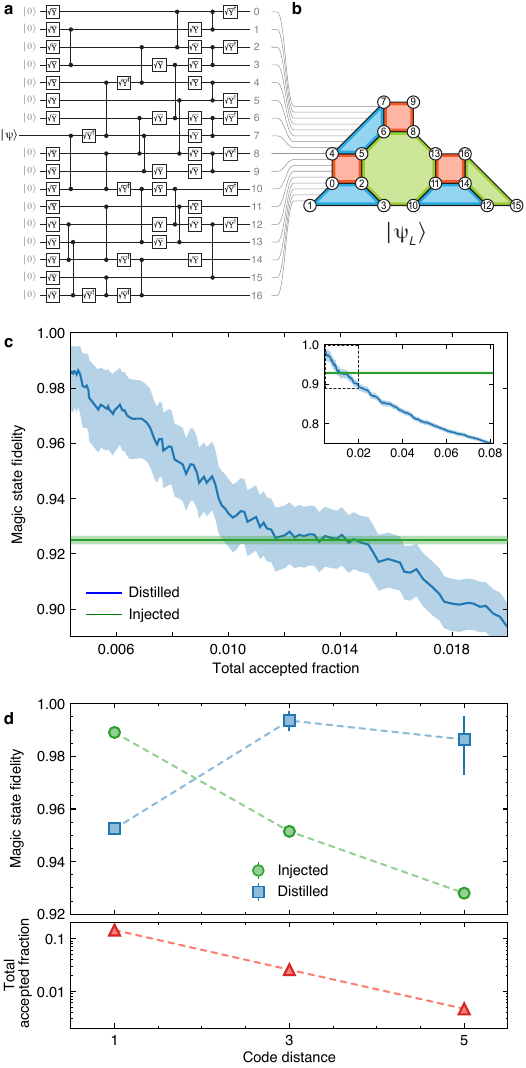}
    \caption{\textbf{Scaling of distillation as a function of data code distance.}
    \textbf{a,}~$d\,{=}\,5$ encoding circuit. 
    \textbf{b,}~$d\,{=}\,5$ color code stabilizers. 
    \textbf{c,}~Output magic state fidelity for $d\,{=}\,5$ distillation (blue) as a function of the total accepted fraction, again showing improvement over the input magic state fidelity (green). Extended range shown in inset. 
    \textbf{d}, Injected (green circles) and distilled (blue squares) magic state fidelity with total acceptance fraction (red), when performing full stabilizer and factory postselection, all as a function of code distance.
    \label{fig:distance}}
\end{figure}

\section*{Extending to larger code distance}
Larger data codes offer stronger protection against physical errors when operated below threshold, and are crucial for scaling to low logical error rates.
To this end, we investigate data codes with larger code distances by performing MSD on five copies of a [[17,1,5]] $d\,{=}\,5$ color code.
After optimizing for our native gate set, we obtain the logical encoding circuit with five entangling layers shown in Fig. \ref{fig:distance}a, with its corresponding stabilizers shown in Fig. \ref{fig:distance}b.
Encoding all five $d\,{=}\,5$ magic states involves 85 physical qubits, which are coherently manipulated in parallel within the entanglement zone of the processor (Methods).
The transversal single- and two-qubit distillation gates and atom moves are exactly the same as in $d\,{=}\,3$.
We apply the same correlated decoding procedure with the MLE decoder and postselection criteria as the $d\,{=}\,3$ case.

The experimental results for the $d\,{=}\,5$ color code are shown in Fig.~\ref{fig:distance}c.
We first note that encoding magic states into larger distance codes results in lower injected fidelity \injfiddfive{}, as the encoding circuit involves more physical gate operations. 
We note that the stabilizer postselection rate required to achieve a comparably high fidelity is lower compared with $d\,{=}\,3$, as our gate fidelities are not yet past the circuit threshold for the data code; this can be improved with further reduction of the physical error rate.
With full stabilizer and factory postselection, we observe a distillation gain of \gainfiddfive{}, from an encoded fidelity of \injfiddfive{} to a distilled fidelity of \distfiddfive{} (Fig. \ref{fig:distance}c).

We compare the MSD performance across code distances in Fig. \ref{fig:distance}d, including physical MSD ($d\,{=}\,1$) and logical MSD with the $d\,{=}\,3$ and $d\,{=}\,5$ color codes.
For physical MSD, preparation of physical magic states is limited only by qubit initialization, measurement and single-qubit gate fidelity.
Without the ability to perform error-correction, physical distillation introduces additional errors, leading to a lower output fidelity.
Shifting to logical qubits, we observe that the injected state fidelity drops as distance increases due to the added errors during the non-fault-tolerant encoding circuit.
However, the data code provides sufficient protection of distillation operations to achieve distillation gain for both $d\,{=}\,3$ and $d\,{=}\,5$.

\section*{Discussion and Outlook}
These experiments demonstrate key ingredients of MSD for universal fault-tolerant quantum computation. 
Leveraging the dynamic reconfigurability and transversal gate operations of the neutral atom platform to realize a logical magic state distillation factory, our approach allows us to probe key aspects of the distillation process.
Such a factory can be combined with mid-circuit measurement and feed-forward~\cite{bluvstein2024logical,graham2023midcircuit,deist2022mid,singh2022mid,lis2023midcircuit,norcia2023midcircuit}, to execute universal quantum algorithms via magic state teleportation.
Although the present experiments clearly demonstrate performance of MSD past the distillation threshold, further improvements in both the fidelity and rate of the MSD factory are required to enable the execution of deep logical circuits.
While at present, the use of higher distance codes results in lower acceptance fraction to achieve large fidelity gain, by improving gate fidelities to values well below the 2D color code threshold, the accepted fraction can remain comparable as the code distance increases, and multiple distillation rounds can be executed for further error suppression.
More specifically, we estimate (Methods) that two-fold reduction in physical error rates will result in distillation gain without stabilizer postselection.

In order to enable efficient 
large-scale universal quantum computation, such fidelity improvements should also come hand-in-hand with further co-design of magic state preparation. 
While magic state distillation represents a foundational approach for implementing non-Clifford operations, and has the advantage of being flexibly adaptable to many data codes, alternative methods with various trade-offs should also be explored.
These include the use of QEC codes with transversal non-Clifford gates~\cite{bombin2015gauge,kubica2015unfolding}, as well as advanced flag protocols~\cite{postler2022demonstration,gupta2024encoding,chamberland2019fault,chamberland2020very,hirano2024leveraging,itogawa2024even} and the recently proposed magic state cultivation~\cite{gidney2024magic} schemes.
Moreover, alternative MSD factories with improved input-to-output ratios or better error suppression~\cite{jones2012multilevel,ogorman2017quantum,litinski2019magic,gidney2019efficient,bravyi2005universal} can be co-designed and explored experimentally within the current framework.
Paving the way towards reliable operation in large-scale quantum computers, our work therefore opens the door for exploration  of hardware efficient generation of quantum magic. 

\section*{Acknowledgements}
We acknowledge helpful discussions with L.~Jiang, M.~Kang, G.~Masella, C.~Pattison, A.~Pi\~{n}eiro~Orioli, Q.~Xu, M.~Yuan, and technical contributions from I.~Paus, C.~Skinker, Q.~Yu.
This work, including the design, assembly and operation of the Gemini-class neutral atom quantum computer was supported by QuEra Computing. Pathfinding work at Harvard and MIT was supported 
by IARPA and the Army Research Office, under the Entangled Logical Qubits program (Cooperative Agreement Number W911NF-23-2-0219), the DARPA ONISQ program (W911NF2010021), 
the DARPA MeasQuIT program (HR0011-24-9-0359), 
the Center for Ultracold Atoms (a NSF Physics Frontiers Center, PHY-1734011), the National Science Foundation (grant number PHY-2012023 and grant number CCF-2313084), the Army Research Office MURI (grant number W911NF-20-1-0082), and QuEra Computing. 

Z.H. acknowledges support from the NSF Graduate Research Fellowship Program under Grant No. 2141064.
D.B. acknowledges support from the NSF Graduate Research Fellowship Program (grant DGE1745303) and The Fannie and John Hertz Foundation. S.J.E. acknowledges support from the National Defense Science and Engineering Graduate (NDSEG) fellowship. T.M. acknowledges support from the Harvard Quantum Initiative Postdoctoral Fellowship in Science and Engineering. M.C. acknowledges support from Department of Energy Computational Science Graduate Fellowship under Award Number DE-SC0020347.

\section*{Author Contributions}
The QuEra team designed, built and ran the experiment and performed data analysis. All authors discussed the results, revised and contributed to the writing of this manuscript.

\section*{Competing interests}
M.G., V.V. and M.D.L. are co-founders and shareholders of QuEra Computing. Authors affiliated with QuEra Computing are employees or interns at QuEra Computing at the time of their contributions.

\newpage

\bibliography{main}
\bibliographystyle{unsrt85}

\include{methods}

\newpage

\setcounter{figure}{0}
\newcounter{EDfig}
\renewcommand{\figurename}{Extended Data Figure}

\begin{figure*}
    \centering
    \includegraphics[width=1.0\linewidth]{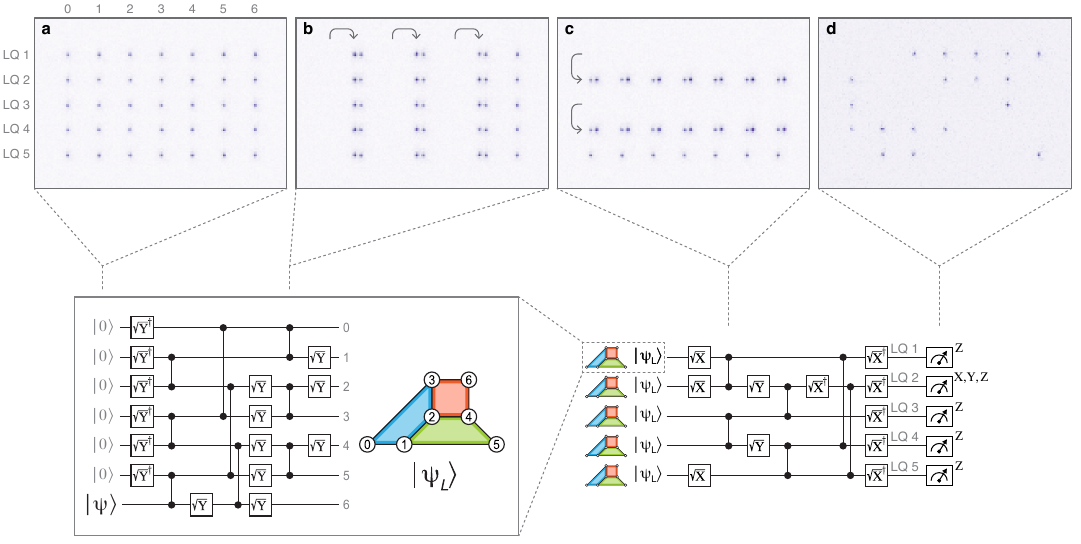}
    \caption{
    \textbf{Experimental layout of magic state distillation factory.}
    \textbf{a,} 
    We arrange 7 to 17 $^{87}$Rb atoms, each corresponding to a physical qubit, into a row. This horizontal register represents a logical qubit, tiled into 5 rows for a total of five logical qubits (LQ1 to LQ5). 
    \textbf{b,} Encoding. Once the register of physical qubits is prepared, we coherently rearrange atoms to perform two-qubit entangling gates using the Rydberg blockade mechanism. We break up the circuit into ``layers'' each containing one set of local rotations, transport, and CZ gates. 
    \textbf{c,} Coherent movement of logical qubits to perform transversal CZ gates. In the case of 5-to-1 distillation, this is achieved in three layers. The circuit as drawn here corresponds 1 to 1 to the atom layout, whereas in Fig.~\ref{fig:d3_distillation} logical qubits LQ1 and LQ2 are swapped for clarity.
    \textbf{d}, Global measurement of qubits after circuit execution. 
    }
    \label{fig:experimental_details}
\end{figure*}

\newpage

\begin{figure*}
    \centering
    \includegraphics[width=1.0\linewidth]{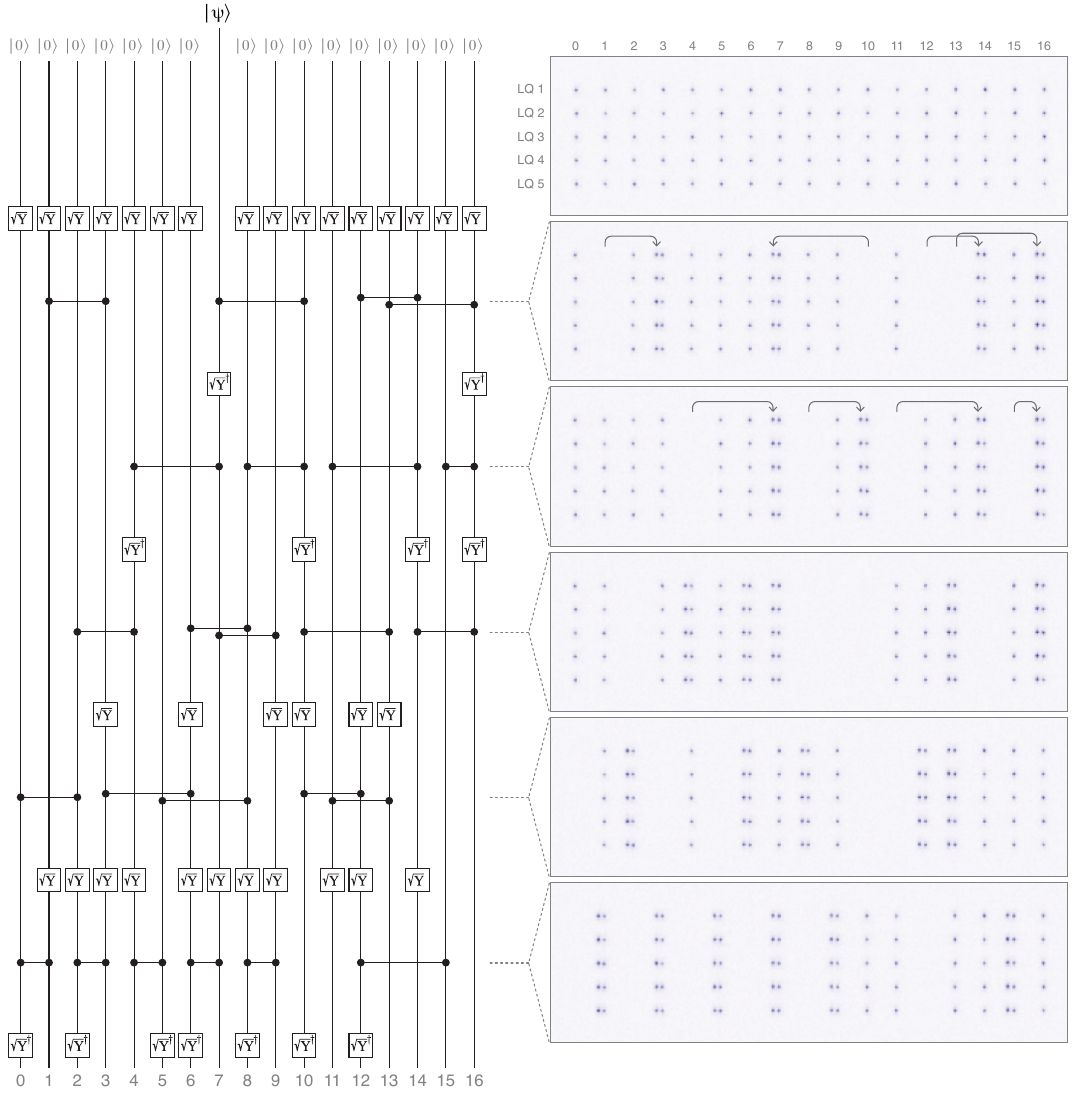}
    \caption{
    \textbf{Experimental layout of $\mathbf{d\,{=}\,5}$ encoding.} The arbitrary-state encoding circuit for the $d\,{=}\,5$ color code (left) is comprised of five entangling gate layers, illustrated by averaged images of the corresponding atom configurations (right), and local gates between the layers.
    We execute encoding with 5x parallelism, one instance per row (LQ1 to LQ5). The horizontal AOD trap array is tiled vertically by the second AOD. For each layer, atoms start in SLM sites, we apply local rotations, pick up and move atoms to their gate location, execute parallel CZ gates, echo (omitted for clarity), and finally move back to SLM sites.
    }
    \label{fig:experimental_details_d5}
\end{figure*}

\begin{figure*}
    \centering
    \includegraphics[width=1.0\linewidth]{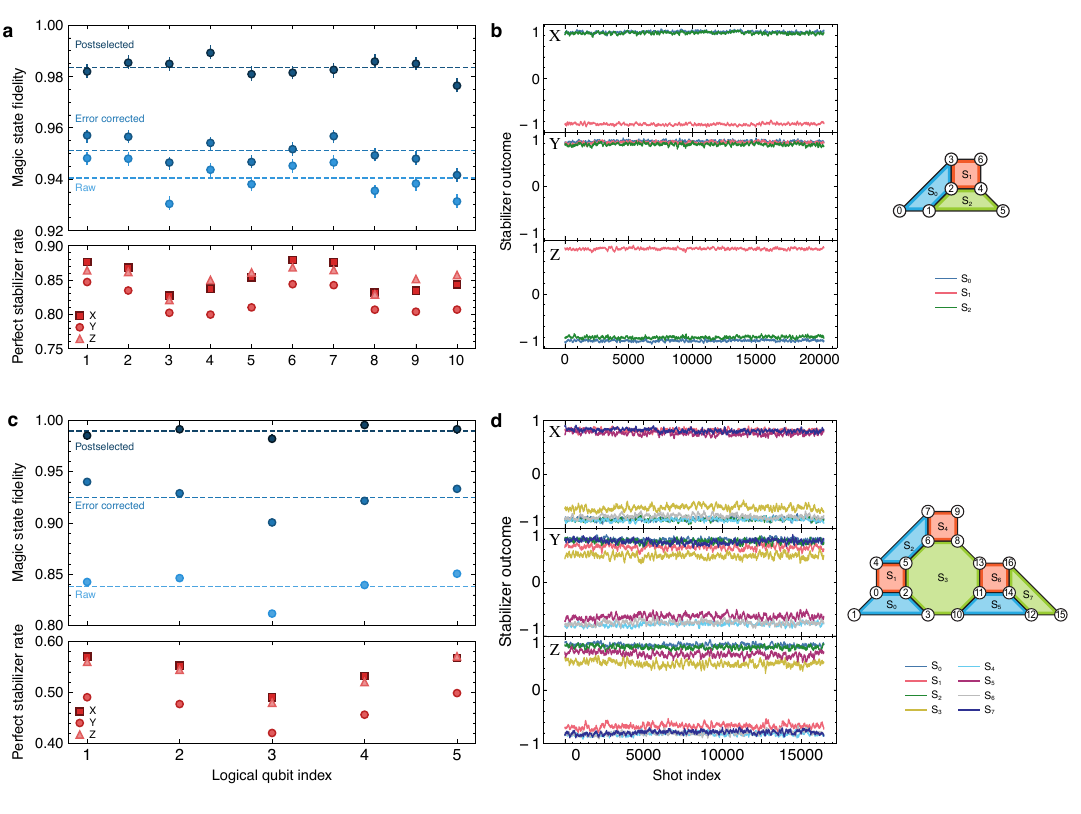}
    \caption{
    \textbf{Encoded magic state fidelity and stabilizers.} 
    \textbf{a,} Spatial dependence of distance-3 magic state encoding fidelity, for the experimental run with no added coherent error. Logical qubits numbered 1-5 and 6-10 are the input qubits to two parallel distillation circuits. We observe some spatial dependence on both the fidelity and perfect stabilizer rate, which we attribute to local single-qubit gate inhomogeneity and two-qubit gate inhomogeneity. 
    \textbf{b,} Time dependence of distance-3 color code stabilizers, for the experimental run with no added coherent error. Time traces are averaged with window size of 100. \textbf{c,} \textbf{d,} Same as \textbf{a} and \textbf{b} for distance-5.
    }
    \label{fig:spatial_time_dep}
\end{figure*}

\begin{figure*}
    \centering
    \includegraphics[width=1.0\linewidth]{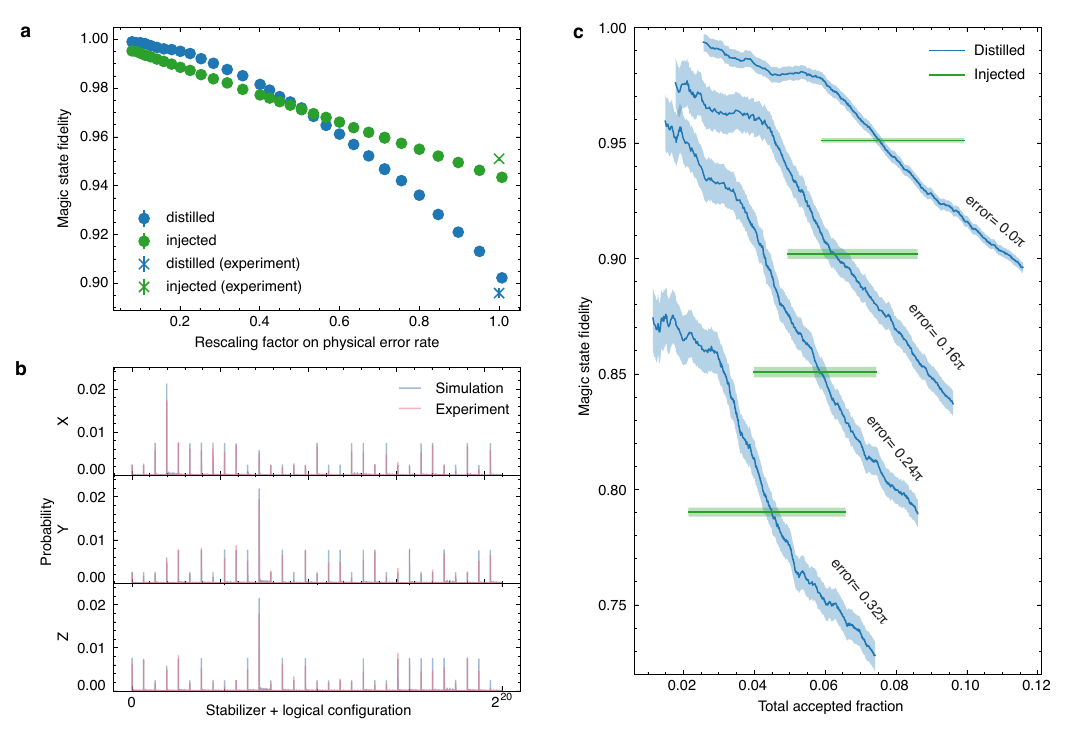}
    \caption{
    \textbf{Additional decoding results.} \textbf{a,} Simulated injected and distilled magic state fidelities as a function of global rescaling of physical error rates, when no stabilizer postselection is applied.
    Relative to our error model for decoding, the physical error rates have been further increased by 1.25$\times$ to match the experimental injected and distilled fidelities.
    \textbf{b,} Simulation and experimental data in table format for $d=3$, sorted into bins corresponding to 3$\times$5=15 stabilizers and 5 logical observables, for a total of $2^{20}$ bins. We see good agreement between simulation and experiment.
    \textbf{c,} Sliding-scale postselection of experimental distillation fidelity with added input errors. Fidelity of the output magic state (blue line) as a function of the total accepted fraction. The accepted fraction range decreases with added errors due to the factory acceptance rate decreasing. Horizontal line segments indicate the error corrected fidelity of the factory input states (green). Shaded regions indicate 68\% confidence intervals.
    }
    \label{fig:decoder_cuts}
\end{figure*}

\newpage

\end{document}

%% file: authors.tex
\author{Pedro Sales Rodriguez}
\thanks{These authors contributed equally.}
\author{John M. Robinson}
\thanks{These authors contributed equally.}
\author{Paul Niklas Jepsen}
\thanks{These authors contributed equally.}
\affiliation{QuEra Computing Inc., Boston, Massachusetts 02135, USA}
\author{Zhiyang He}
\affiliation{QuEra Computing Inc., Boston, Massachusetts 02135, USA}
\affiliation{Department of Mathematics, Massachusetts Institute of Technology, Cambridge, Massachusetts 02138, USA.}
\author{Casey Duckering}
\author{Chen Zhao}
\author{Kai-Hsin Wu}
\author{Joseph Campo}
\author{Kevin Bagnall}
\author{Minho Kwon}
\author{Thomas Karolyshyn}
\author{Phillip Weinberg}
\affiliation{QuEra Computing Inc., Boston, Massachusetts 02135, USA}
\author{Madelyn Cain}
\author{Simon J. Evered}
\author{Alexandra A. Geim}
\author{Marcin Kalinowski}
\author{Sophie H. Li}
\author{Tom Manovitz}
\affiliation{Department of Physics, Harvard University, Cambridge, Massachusetts 02138, USA.}
\author{Jesse Amato-Grill}
\author{James I. Basham}
\author{Liane Bernstein}
\author{Boris Braverman}
\author{Alexei Bylinskii}
\author{Adam Choukri}
\author{Robert DeAngelo}
\author{Fang Fang}
\author{Connor Fieweger}
\author{Paige Frederick}
\author{David Haines}
\author{Majd Hamdan}
\author{Julian Hammett}
\author{Ning Hsu}
\author{Ming-Guang Hu}
\author{Florian Huber}
\author{Ningyuan Jia}
\author{Dhruv Kedar}
\author{Milan Kornja\v{c}a}
\author{Fangli Liu}
\author{John Long}
\author{Jonathan Lopatin}
\author{Pedro~L.~S.~Lopes}
\author{Xiu-Zhe Luo}
\author{Tommaso Macr\`i}
\author{Ognjen Markovi\'c}
\author{Luis A. Mart\'inez-Mart\'inez}
\author{Xianmei Meng}
\author{Stefan Ostermann}
\author{Evgeny Ostroumov}
\author{David Paquette}
\author{Zexuan Qiang}
\author{Vadim Shofman}
\author{Anshuman Singh}
\author{Manuj Singh}
\author{Nandan Sinha}
\author{Henry Thoreen}
\author{Noel Wan}
\author{Yiping Wang}
\author{Daniel Waxman-Lenz}
\author{Tak Wong}
\author{Jonathan Wurtz}
\author{Andrii Zhdanov}
\author{Laurent Zheng}
\affiliation{QuEra Computing Inc., Boston, Massachusetts 02135, USA}
\author{Markus Greiner}
\affiliation{Department of Physics, Harvard University, Cambridge, Massachusetts 02138, USA.}
\author{Alexander Keesling}
\author{Nathan Gemelke}
\affiliation{QuEra Computing Inc., Boston, Massachusetts 02135, USA}
\author{Vladan Vuleti\'c}
\affiliation{Department of Physics, Massachusetts Institute of Technology, Cambridge, Massachusetts 02138, USA.}
\author{Takuya Kitagawa}
\author{Sheng-Tao Wang}
\affiliation{QuEra Computing Inc., Boston, Massachusetts 02135, USA}
\author{Dolev Bluvstein}
\author{Mikhail D. Lukin}
\affiliation{Department of Physics, Harvard University, Cambridge, Massachusetts 02138, USA.}
\author{Alexander Lukin}
\author{Hengyun Zhou}
\email{hyzhou@quera.com}
\author{Sergio H. Cant\'u} 
\email{scantu@quera.com}
\affiliation{QuEra Computing Inc., Boston, Massachusetts 02135, USA}

%% file: methods.tex
\section*{Methods}

\newcommand{\methodssubsection}[1]{\vspace{1em}{\noindent\textbf{#1}}\vspace{0.25em}

}
\newcommand{\edfig}{Extended Data Fig.}

\methodssubsection{System overview}
\noindent All experiments described in this work were performed on QuEra's Gemini-class neutral atom quantum computer. More detailed information and characterization will be described in an upcoming manuscript~\cite{robinson2024benchmarking}. The system is
based on neutral $^{87}$Rb atoms trapped in reconfigurable optical tweezers~\cite{barredo2016atom,endres2016atom,kim2016in,ebadi2021quantum}.
We laser cool and trap atoms in a magneto-optical-trap, and subsequently load into a fixed set of optical tweezers generated by a spatial light modulator (SLM).
Atoms are then coherently rearranged by dynamic tweezers at 852 nm generated by a crossed pair of acoustic-optic deflectors (AODs).

Qubits are encoded in the $m_F$=0 hyperfine ground states, $\ket{0}\equiv\ket{F\!=\!1,m_f\!=\!0}$ and $\ket{1}\equiv\ket{F\!=\!2,m_f\!=\!0}$, with a $T_2$ coherence time of approximately 2~s \cite{robinson2024benchmarking}.
Single-qubit gates are performed via Raman transitions~\cite{levine2022dispersive}, with a laser red-detuned from $5P_{1/2}$ by 350~GHz. We drive global single-qubit rotations at 650~kHz by illuminating the entire array along the quantization axis, and locally at 250~kHz by addressing atoms using another pair of AODs~\cite{bluvstein2024logical}.
Two-qubit gates are mediated by Rydberg interactions, which we achieve by driving atoms in $\ket{1}$  to $53\text{S}_{1/2}$ in a two-photon process via $6\text{P}_{3/2}$ with intermediate state detuning of 6 GHz~\cite{levine2019parallel,evered2023high,ma2023high,jandura2022time,scholl2023erasure}. We perform readout globally, by heating and ejecting atoms in $\ket{1}$ with resonant light, followed by fluorescence imaging of the remaining atoms.

\methodssubsection{Circuit details and calibration}
\noindent We deterministically load and prepare the atoms into a rectangular grid of $17\times 5$ SLM traps.
The same regular grid of SLM sites is used to run both the $d\,{=}\,3$ and $d\,{=}\,5$ distillation experiments (Extended Data Fig. \ref{fig:experimental_details} and \ref{fig:experimental_details_d5}).
During circuit execution, atoms are rearranged entirely within the 5-row-wide ``entanglement zone'', and illuminated with 1013 nm and 420 nm light that couples qubit state $\ket{1}$ to the Rydberg state. 
To execute CZ gates, we coherently move atoms such that gate pairs are 2~$\mu$m away, within the Rydberg blockade radius, while keeping 8~$\mu$m separation between independent gate pairs.
We perform parallel horizontal moves during data code encoding to prepare a logical qubit in each row. 
Once encoded, we move the rows of logical qubits using parallel vertical moves.
To preserve qubit coherence, all moves are accompanied by dynamical decoupling implemented with global single-qubit pulses.
Local single-qubit gates around the X-Y plane are executed in between CZ layers and we echo the induced local light shifts using a global single-qubit gate pulse.

We now detail changes in logical operator and stabilizer conventions due to the circuit optimizations that we apply. 
In both $d\!=\!3$ and $d\!=\!5$ encoding, we physically implement the first layer of $\sqrt{Y}$ gates with a global pulse.
We also substitute local $\sqrt{Y}^\dagger$ with $\sqrt{Y}$ to improve parallelism.
This changes the basis of the physical input, so to inject a logical (1,1,1) we physically prepare the (-1,-1,-1) state.
We further optimize by pre-applying the first set of $\sqrt{X}$ gates required for distillation on the physical qubits prior to encoding: qubits 1, 2 and 5 are prepared into (-1,1,-1), which after encoding becomes (1,-1,1).
Finally, single qubit echoes during the encoding and distillation circuit redefine the color code stabilizer basis.
We classically track this through encoding and distillation to recover the original stabilizer basis. 
Echoes during distillation also flip the distillation syndrome outcomes.
This means that our actual acceptance case is 0011, different from the acceptance case of 1011 from the circuit in Fig.~\ref{fig:d3_distillation}a.

We use quantum state tomography to evaluate the fidelity of logical magic states.
Measurement of the injected state fidelity is done by applying a global tomography pulse to all qubits and subsequently measuring all three bases, see \edfig{}~\ref{fig:spatial_time_dep} for examples.
To measure the fidelity after distillation, we apply transversal single qubit gates to the output logical qubit to sample the three bases.
In this way, the other four logical qubits are always measured in the Z basis.
Loss during the circuit can lead to a biased error in the magic state fidelity.
To mitigate this, we interleave measurements in all basis states: $\pm$X, $\pm$Y, $\pm$Z and average the results. 
We track the injected fidelity by interleaving one shot of magic state injection with no distillation for every seven shots of full factory execution.
This protects against bias due to potential systematic drift during data taking.
Throughout the run we monitor the perfect stabilizer rate of the injection circuit as a proxy for gate calibrations (Extended Data Fig.~\ref{fig:spatial_time_dep}b,d).

A single factory instance using $d\!=\!3$ and $d\!=\!5$ data codes requires 35 and 85 atoms, respectively.
To improve data rates, we run \textit{two} independent parallel instances of the $d\!=\!3$ factory, requiring a total of 70 atoms.
We ran $658,562$ shots of the $d\!=\!3$ experiment, split into a total of $2\!\times\!576,131\!=\!1,152,262$ factory runs and $10\!\times\!82,431\!=\!824,310$ encoding tomography runs. Added error datasets consist of 143,000 shots for each added error, corresponding to 251,000 factory runs and 175,000 encoding tomography runs. For $d\!=\!5$, we ran a total of 259,261 factory shots and $5\!\times37,108\!=\!185,540$ encoding tomography shots. Shots are evenly split into X,Y,Z basis for all experiments.

\methodssubsection{Error model} 
\noindent We use randomized benchmarking (RB) to calibrate and benchmark single- and two-qubit gates. We measure a global (local) amplitude robust single-qubit gate fidelity of 99.978(1)\% (99.81(2)\%). 
We benchmark controlled-phase (CZ) two-qubit gates by driving pairs of blockaded atoms with alternating single-qubit and two-qubit gates \cite{evered2023high}. We measure the return probability to $\ket{00}$ as a function of the number of entangling gates, resulting in a fidelity of 99.42(1)\% per entangling gate. We estimate state preparation and measurement errors to be a total of 1\%.

To simulate the impact of various error sources on the circuits, we model the error sources as depolarizing Pauli channels.
Errors due to global and local single-qubit gates are incorporated as single-qubit channels.
Two-qubit gate error is modeled by a two-qubit depolarizing Pauli channel, biased towards Z and ZZ phase flip channels.
Movement of atoms via AOD also induces errors but in two distinct ways. On the moving atoms, tweezer light induces a qubit frequency shift, resulting in Z errors. During the move duration we also account for the idling errors on all qubits. For each error type, we assume uniform error across the atom array and the magnitude of errors is derived from independent benchmarking of each operation.
Overall, our error model shows good agreement with the experimentally observed stabilizer and logical outcomes (\edfig{}~\ref{fig:decoder_cuts}b).

\methodssubsection{Probing distillation code error suppression}

5-to-1 MSD achieves a quadratic suppression in infidelity of the input magic states. We probe this phenomenon in experiment and in (noiseless) simulation by applying coherent $Z$ errors to input magic states and recording output magic state fidelity (Fig.~\ref{fig:d3_distillation}c) as well as factory acceptance rate (Fig.~\ref{fig:d3_distillation}d). 

In experiments, we apply coherent Z errors of $0.32\pi$, $0.24\pi$, $0.16\pi$ and $0$ to the five physical magic states, which are then injected into $d=3$ color codes and distilled. 
The output fidelity is plotted against the injected postselected fidelity in Fig.~\ref{fig:d3_distillation}c.
The distillation output fidelity performs only error correction on the output logical qubit, but postselects on perfect stabilizers on the distillation syndrome logical qubits.
The injected postselected fidelity, on the other hand, postselects on perfect stabilizers on the target logical qubit itself.
We choose this comparison to more clearly highlight the distillation behavior of the logical information.
We see that all four data points show distillation gain; namely, the error-corrected output fidelity is higher than the postselected injected fidelity.
In \edfig~\ref{fig:decoder_cuts}c, we further show the results for different distillation stabilizer postselection thresholds.

We also numerically simulate the performance of the ideal distillation circuit subject to coherent input errors.
The output fidelity is plotted against input physical magic state fidelity, calculated based on the applied error (gray curve, Fig.~\ref{fig:d3_distillation}c). 
We observe the expected quadratic suppression in input infidelity, which is in good agreement with our experimental data (blue points).
Note that when the input fidelity is at 0.80, which is lower than the frequently-quoted 5-to-1 MSD distillation threshold of 83\%~\cite{bravyi2005universal}, we still observe an improved output fidelity.
This is because the usual distillation threshold is computed for incoherent errors, while the threshold for our applied coherent errors is lower.
We also observe good agreement with experimental data for the factory acceptance rate (Fig.~\ref{fig:d3_distillation}d). 
Overall, our experimental data closely aligns with the theoretical predictions of 5-to-1 MSD.

\methodssubsection{Comparison with alternative methods for magic state preparation}

\noindent In this section, we compare different methods for magic state preparation, including alternative injection or projection-based schemes, and other magic state distillation factories.

There are a few natural approaches to prepare logical magic states.
One could use an (often non-fault-tolerant) encoding circuit, measure the data code stabilizers to project into the target logical state, or measure certain operators for which the target logical magic state is an eigenstate.
Some of these operations can further serve to detect errors, in order to boost the fidelity of the resulting magic state.
These protocols can be further expanded with flag qubits or extended with error-correction cycles for improved fidelity. 

Prior experiments on trapped ions~\cite{postler2022demonstration} and superconducting qubits~\cite{gupta2024encoding,ye2023logical} have demonstrated different combinations of these techniques.
This includes magic state preparation based on unitary encoding~\cite{postler2022demonstration} and stabilizer measurement projection~\cite{ye2023logical}, as well as the further use of flagged verification schemes and error detection/correction to achieve fault tolerance against any single physical error~\cite{gupta2024encoding,postler2022demonstration}.

The above techniques of magic state injection and verification can produce magic states with fairly high fidelity, which could serve as the input into MSD factories in the future~\cite{hirano2024leveraging}.
However, these techniques have some noteworthy limitations due to their direct use of physical operations, in contrast to protected logical operations of the data code as in MSD.
First, direct injection of physical magic states without further verification will have a performance limited by the physical magic state fidelity, which is insufficient for large-scale quantum computing.
Second, operating the verification protocols at higher distances or higher physical error rates generally increases the complexity of ancilla preparation and/or post-selection overhead significantly~\cite{chamberland2019fault,chamberland2020very,gidney2024magic}.

For these reasons, protocols that make use of an inner data code to protect operations, such as MSD, are a crucial primitive as we scale to lower logical error rates.
Existing implementations of MSD achieve this with physical qubits~\cite{souza2011experimental,brown2023advances}, which do not provide protection of Clifford operations within the MSD factory.
Thus, our demonstration of logical MSD with an inner data code is a crucial step towards further improvements of magic state preparation.

Our experiment focuses on the implementation of 5-to-1 MSD~\cite{bravyi2005universal}, as it exemplifies the principles of MSD with relatively low resource requirements.
It has the downside that with perfect distillation code operations, the factory only has a $1/6$ acceptance rate, and it only achieves quadratic suppression despite the code being distance 3.
For future, large-scale operation, it may instead be desirable to employ MSD factories with higher distillation rate, better error suppression scaling, and which have unity acceptance rate in the absence of input errors~\cite{jones2012multilevel,ogorman2017quantum,litinski2019magic,gidney2019efficient,bravyi2005universal}.
However, some of the key ingredients we demonstrated, such as the use of parallel operations via transversal gates and sliding scale post-selection based on stabilizer readouts, are likely broadly useful for future experiments.

\methodssubsection{Design and optimization of state injection circuits}

\noindent To achieve high fidelities, we optimize the implementation of several key quantum circuits. In this section, we focus on arbitrary state injection circuits for the $d\!=\!3$ and $d\!=\!5$ color codes, while the next section will discuss optimizations of logical MSD circuits.
We primarily focus on reducing the number of entangling gate layers, since the gate infidelity and associated move errors are a significant contributor to our error budget.
To the best of our knowledge, previous unitary injection circuits for the $d\!=\!3$ color code require 4 entangling gate layers~\cite{postler2022demonstration,goto2016minimizing,mayer2024benchmarking}.
While exhaustive search over all possible 7-qubit, 3-layer injection circuits is sufficient for $d\!=\!3$, we develop more efficient methods for $d\!=\!5$ to find low-depth circuits with good atom layouts.

We present an algorithm based on matrix row reduction~\cite{gottesman1997stabilizer,cleve1997efficient} to find an injection circuit for 2D color codes.
Simple extensions to this algorithm may work well for any CSS code in general.
Executing the algorithm gives \textit{some} injection circuit, which is unlikely to be optimal.

The algorithm operates on a matrix representation of the checks and logical operator of the code.
Each row corresponds to a data qubit of the code and each column is either a check or a logical operator.
The matrix entry $M_{qc}$ is $1$ if the check or logical $c$ contains qubit $q$, and is $0$ otherwise.
Performing row operations (adding one row to another) on this matrix corresponds to the application of CNOTs, while performing column operations (adding one column to another) corresponds to redefining stabilizers and logical operators.
We find the circuits shown in Fig.~\ref{fig:d3_colorcode}a and Fig.~\ref{fig:distance}a using the following simple heuristics and search methods:
\begin{enumerate}
    \item Choose row operations in \textit{layers} where we pick the ``best'' $n/2$ disjoint pairs of rows for row operations before reusing rows in the next layer.  This maximizes circuit parallelism because each row operation will become a CNOT.
    \item Every row operation ideally reduces (or sometimes maintains) the total number of 1 entries.
    \item A row operation is preferred if it leaves the updated row more similar (by Hamming distance) to another row.  This enables a future row operation to be more effective.
    \item Prioritize row operations to remove 1 entries from high-weight columns (high relative numbers of 1 entries).  If certain columns are very high-weight near row-reduction completion, backtrack and prioritize them sooner.
    \item While backtracking to try new choices, prioritize minimizing number of operation layers over number of operations.
    \item Column operations do not need to be optimal as they do not impact the circuit, they only redefine the stabilizer basis.
\end{enumerate}

For example, the $d\,{=}\,3$ color code depicted in Fig.~\ref{fig:d3_colorcode}b has checks $S_0\,{=}\,Z_0Z_1Z_2Z_3$, $S_1\,{=}\,Z_1Z_2Z_4Z_5$, $S_2\,{=}\,Z_2Z_3Z_4Z_6$, $S_3\,{=}\,X_0X_1X_2X_3$, $S_4\,{=}\,X_1X_2X_4X_5$, $S_5\,{=}\,X_2X_3X_4X_6$ and logical operators $L_{Z0}\,{=}\,Z_0Z_1Z_5$, $L_{X0}\,{=}\,X_0X_1X_5$.
Due to the self-dual structure where X and Z checks match, this can be represented with the matrix
$$
M^0 ~=~~ \begin{matrix}
    & S_0 & S_1 & S_2 & L_0 \\
q_0 & 1 & - & - & 1 \\
q_1 & 1 & 1 & - & 1 \\
q_2 & 1 & 1 & 1 & - \\
q_3 & 1 & - & 1 & - \\
q_4 & - & 1 & 1 & - \\
q_5 & - & 1 & - & 1 \\
q_6 & - & - & 1 & - \\
\end{matrix}
$$
where zero is shown as ``$-$'' for visual clarity.

Our goal is to find a sequence of row and column operations via matrix row reduction under addition modulo 2.
For example, the row operation $0 \rightarrow 2$ results in 
$$
\begin{matrix}
    & S_0 & S_1 & S_2 & L_0 \\
\boldsymbol{q_0} & \boldsymbol{1} & \boldsymbol{-} & \boldsymbol{-} & \boldsymbol{1} \\
q_1 & 1 & 1 & - & 1 \\
\boldsymbol{q_2} & \boldsymbol{-} & \boldsymbol{1} & \boldsymbol{1} & \boldsymbol{1} \\
q_3 & 1 & - & 1 & - \\
q_4 & - & 1 & 1 & - \\
q_5 & - & 1 & - & 1 \\
q_6 & - & - & 1 & - \\
\end{matrix}
$$
Note that columns corresponding to the logical operator(s) must not be source columns (e.g. $L_i \rightarrow \circ$ is not allowed), but may be target columns.

We find the best row operations 
$R_{ops} = [
0 \rightarrow 1,
3 \rightarrow 2,
5 \rightarrow 4,
0 \rightarrow 3,
2 \rightarrow 5,
4 \rightarrow 6,
2 \rightarrow 1,
4 \rightarrow 3,
6 \rightarrow 5]
$
and column operations
$C_{ops} = [S_0 \rightarrow L_{Z0},
S_2 \rightarrow L_{Z0}]$
result in our final matrix $M^{final}$
$$
M^{final} ~= ~~\begin{matrix}
    & S_0 & S_1 & S_2 & L_0 \\
q_0 & 1 & - & - & - \\
q_1 & - & - & - & - \\
q_2 & - & 1 & - & - \\
q_3 & - & - & - & - \\
q_4 & - & - & 1 & - \\
q_5 & - & - & - & - \\
q_6 & - & - & - & 1 \\
\end{matrix}
$$

This solution explicitly defines our encoding circuit:
\begin{enumerate}
    \item On the qubit where $M^{final}_{q_i, L_{Zj}} = 1$, prepare the injected state $\ket\psi$.
    \item For other $q_i$ where $M^{final}_{q_i, S_j} = 1$ for some $S_j$, prepare $q_i$ in the $\ket+$ state.
    \item Prepare all other qubits in the $\ket0$ state.
    \item For each entry $s \rightarrow t$ of $R_{ops}$, \textit{in reverse order} add a CNOT gate with control $q_s$ and target $q_t$.  $C_{ops}$ does not impact the circuit.
    \item Use circuit identities to convert to hardware-supported gates:
    \begin{enumerate}
        \item Preparing a qubit in the $\ket+$ state becomes preparing in the $\ket0$ state followed by $\sqrt{Y}$.
        \item $\mathrm{CNOT}_{ij}$ becomes $\sqrt{Y_j}^\dagger$, $\mathrm{CZ}_{ij}$, and $\sqrt{Y_j}$.
        \item Adjacent $\sqrt{Y_j}$ and $\sqrt{Y_j}^\dagger$ cancel.
    \end{enumerate}
\end{enumerate}

The resulting circuit non-fault-tolerantly prepares the logical code state $\ket{\overline{\psi}}$.
For the $d\!=\!3$ color code, the solution above gives the 9-gate, 3-layer encoding circuit.
For the [[17, 1, 5]] color code, the best solution we find has five layers and 24 CNOT/CZ gates (Fig.~\ref{fig:distance}a) with
$R_{ops} = [
1  \rightarrow 0,
3  \rightarrow 2,
4  \rightarrow 5,
7  \rightarrow 6,
9  \rightarrow 8,
15 \rightarrow 12,
2  \rightarrow 0,
6  \rightarrow 3,
8  \rightarrow 5,
12 \rightarrow 10,
13 \rightarrow 11,
2  \rightarrow 4,
8  \rightarrow 6,
9  \rightarrow 7,
10 \rightarrow 13,
16 \rightarrow 14,
4  \rightarrow 7,
8  \rightarrow 10,
14 \rightarrow 11,
15 \rightarrow 16,
3  \rightarrow 1,
7  \rightarrow 10,
14 \rightarrow 12,
16 \rightarrow 13
]$.

\methodssubsection{Design and optimization of distillation circuit}

\noindent The 5-to-1 distillation protocol consists of running the unencoding circuit of the $[[5, 1, 3]]$ perfect code, followed by measurements of the four logical qubits that correspond to stabilizers of the distillation code.
To implement this protocol, we start with an unencoding circuit with low entangling gate count~\cite{gidney2024tweet}, previously optimized from Ref.~\cite{laflamme1996perfect}, and further optimize it for our hardware.
Our optimizations aim to reduce the number of local single-qubit gates as well as the number of entangling gate layers, since these have larger contributions to the infidelity. 
We utilize a variety of techniques to achieve this:
\begin{enumerate}[nolistsep]
    \item Reordering of qubits and commuting gates. The final circuit includes three rounds of CZ gates separated by local single-qubit gates.
    \item Use of circuit identities, such as $\frac{1+i}{\sqrt{2}}H = X^{1/2}SX^{1/2}$.
    \item Absorbing certain operations into the initial state or measurement, without changing the ideal initial state, the post-selection basis, or affecting the quadratic error suppression of the distillation circuit.
\end{enumerate}

Our optimized circuit is shown in Fig.~\ref{fig:d3_distillation}a.
Note that in standard 5-to-1 MSD, the successful distillation syndrome is 0000, as depicted in Fig.~\ref{fig:architecture}. Our optimizations flipped it to 1011.
As we used identities related to the initial state inputs and final post-selection, this circuit is an unencoding circuit of a five-qubit code which is equivalent to the perfect code up to Clifford operations. 
For the purpose of magic state distillation, it achieves quadratic suppression in infidelity.

\methodssubsection{Design and optimization of atom layout}

\noindent The optimized circuits described in the previous two sections need efficient implementations of atom movement.
Here, we describe our design process for finding circuit-specialized atom move sequences.

We design logical circuits with transversal operations to have a 2D product structure, where transversal operations are horizontally parallel and logical state injection is vertically parallel (Fig.~\ref{fig:architecture}).
Thus, we lay out each logical qubit linearly in the same row.
All atoms have a home position in a static SLM trap, and for each layer of gates, we pick fewer than half the atoms, move them horizontally or vertically near their gate partners, and move them back.
In order to minimize atom transfer, we optimize for an atom ordering and circuit layers where \textit{none of the moves reorder atoms} and the \textit{move distances are minimized}.
An atom order is valid for given circuit layers if
$$
\begin{matrix}
\mathrm{max}(i, j) < \mathrm{max}(k, l)
\\
\forall~ \mathrm{CZ}_{ij}, \mathrm{CZ}_{kl} \in \mathrm{Layer} ~|~i<k
\\
\forall~ \mathrm{Layer} \in \mathrm{Circuit}
\end{matrix}
$$
where a layer is a set of CZ gates that may be executed in parallel without changing the meaning of the circuit.

We use a combination of hand-optimization over choice of circuit layers and brute-force search over atom orders.
The index numbers labeled in the encoding circuit show these optimal qubit orders.
The order of the five (logical) qubits in Fig.~\ref{fig:d3_distillation}a has the first two qubits swapped.

\methodssubsection{Approach to simulation of magic state distillation circuit performance}

\noindent Our full circuit, which injects five physical magic states into five logical magic states in the color code, and then performs logical magic state distillation, is supported on $35$ qubits in the $d\,{=}\,3$ case and $85$ qubits in the $d\,{=}\,5$ case.
The injection and distillation circuits are entirely Clifford, with the non-Clifford-ness coming only from the input states. 
This poses a challenge towards utilizing standard simulation methods. 
Since the input states are magic states, standard Clifford circuit simulation tools such as Stim~\cite{gidney2021stim} cannot be applied directly. 
The circuit size of 85 means state-vector simulation is intractable, and approximation methods such as matrix product states simulations become technically and computationally consuming.
While methods such as extended Clifford simulation could be used~\cite{bravyi2019simulation}, existing open-source implementations only support up to 64 physical qubits~\cite{javadi-abhari2024quantum}.
For these reasons, we developed a simulation technique which we refer to as \textit{Input-Decoupled Noise Learning}~\cite{he2025channel}, where learning of the noise channel is separated from simulating the actual state of the logical circuit.

The key idea of our approach is that the analysis of noise can largely be separated from the analysis of the ideal logical action itself.
The ideal logical circuit can be viewed as a channel that maps some input quantum state to classical bit strings, $\mathcal{C}: \left(\mathbb{C}^2\right)^{\otimes 5}\rightarrow \mathbb{F}_2^5$, where the bit strings correspond to logical measurement outcomes.
Since this is an ideal logical circuit involving only five qubits, it can be readily simulated.
Under a Pauli noise model $\sigma$ and a Clifford circuit $C$, the combined effect of noise and error correction is to apply additional logical Pauli operations, which further map the logical outcomes $\mathcal{E}_\sigma:\mathbb{F}_2^5\rightarrow \mathbb{F}_2^5$.
Since this combined effect involves only Pauli operators and Clifford circuits, we can efficiently simulate it via error sampling and performing decoding.
More generally, the same approach can be applied whenever error sampling and decoding can be done efficiently.
The full simulation result can then be obtained by composing the two channels $\mathcal{E}_\sigma\circ\mathcal{C}$.

Learning the channel $\mathcal{E}_\sigma$ is implemented as follows. 
We use Stim~\cite{gidney2021stim} to simulate the noisy logical circuit but we replace the input physical magic states by a special 5-qubit entangled state chosen to make the logical measurement results deterministic.
This special 5-qubit state is generated by running the noise-free inverse of the logical circuit.
When measuring the output logical in the X, Y, or Z basis, the state-prep for the special state begins by preparing $\ket{+_L}$, $\ket{+i_L}$ or $\ket{0_L}$ of the 5-qubit distillation code, respectively.
This ensures that the measurement results on all five logical qubits are $+1$ in the absence of errors, and that the simulation is fully Clifford and therefore efficient.
We perform decoding based on the simulated syndromes $\tilde{s}$ (see following sections for details of our decoder), resulting in the final logical measurement result $l\in \mathbb{F}_2^5$ characterizing logical flips caused by circuit noise.
The decoding is done either using the syndrome information of the four logical qubits of factory post-selection, or using syndrome information of all five logical qubits during tomography, resulting in the appropriate channel in each case.
With a large amount of samples, which can be efficiently generated, we can learn the classical logical error channel $\mathcal{E}_\sigma$ to high accuracy. 

For our logical circuit involving five logical qubits, we can easily obtain the ideal logical circuit channel $\mathcal{C}$.
Using Qiskit~\cite{javadi-abhari2024quantum}, we implement the ideal distillation circuit with magic state inputs.
Note that this simulation also supports noise applied on the magic state input.
We directly calculate the 5-bit logical output for this circuit, producing the channel $\mathcal{C}$.
The final output magic state fidelity can then be computed by composing the channels $\mathcal{E}_\sigma\circ\mathcal{C}$.

With this approach, for a Clifford physical circuit with non-Pauli inputs, our method decouples the Pauli noise in the physical circuit from the input states, and learns the noise-induced logical errors efficiently.
Beyond this example, we expect our techniques to have further applications as we scale to larger quantum codes and more complex logical circuits.

\methodssubsection{Estimation of confidence intervals}

\noindent When performing quantum state tomography to estimate the logical fidelity, it is possible that the reconstructed density matrix is not positive semi-definite, causing the calculated fidelity confidence interval to exceed 1~\cite{granade2017practical}.
To address this and obtain meaningful confidence intervals, we use Bayesian analysis to calculate posterior probabilities~\cite{rice2007mathematical}.

Consider quantum state tomography, with $\vec{n}=(n_x, n_y, n_z)$ measurements in the X, Y, Z basis, respectively.
Denote the number of $\ket{0}$ outcomes decoded as $\vec{m}=(m_x, m_y, m_z)$.
We would like to extract the probability distribution of true fidelity values $\mathcal{F}$ that could produce these measurement results.
To this end, we apply Bayes' rule:
\begin{align}
P(\mathcal{F}=F|\vec{m},\vec{n})=\frac{P(\vec{m},\vec{n}|\mathcal{F}=F)P_{prior}(\mathcal{F}=F)}{P(\vec{m},\vec{n})},
\label{eq:bayes}
\end{align}
where $\vec{m},\vec{n}$ denote the observed measurement outcomes and $P_{prior}$ is the prior distribution.
We assume a prior distribution of density matrices that has a uniform random distribution within the Bloch sphere, and expand the right hand side of Eq.~(\ref{eq:bayes}) over quantum mixed states.
Note that this prior has lower weight on the state with unity fidelity, and is therefore more conservative than usual fidelity calculations that correspond to a uniform prior over fidelity values, instead of over the Bloch sphere. In the limit of a large number of samples, the difference between different priors will have a negligible effect.

We (numerically) compute this distribution over the Bloch sphere via
$$
P(\vec{m},\vec{n}|\mathcal{F}=F)
= \int_{\vec{v}|\mathbf{F}_{\ket{SH}}(\vec{v})=F} P(\vec{\mathcal V}=\vec{m},\vec{n}|\vec{v})d\vec{v}
$$
$$
P_{prior}(\mathcal{F}=F)
= \int_{\vec{v}|\mathbf{F}_{\ket{SH}}(\vec{v})=F} 1d\vec{v}
$$
where $\vec{v}=(x, y, z)$ is the Bloch sphere vector representation of the density matrix, 
\begin{equation}
\label{eq:fidelity}
\mathbf{F}_{\ket{SH}}(\vec{v})=\frac{1}{2} + \frac{x+y+z}{2\sqrt3}
\end{equation}
is the fidelity of the mixed state $\vec{v}$ relative to our desired $\ket{SH}$ magic state $\vec{v}_{\ket{SH}}=\frac{1}{\sqrt3}(1,1,1)$ and $P(\vec{m},\vec{n})$ is a normalization constant.
Intuitively, we integrate over all mixed states with the same fidelity.

\methodssubsection{Decoding and post-selection methods}

\noindent At the end of our MSD protocol, we transversally measure all physical qubits of the four distillation syndrome logical qubits in the Z basis, and all physical qubits of the output magic state in one of the X, Y or Z basis for logical tomography.
We utilize two decoding methods for our data: a maximum likelihood decoder (MLD) constructed by direct sampling of a look-up table, and a most likely error (MLE) decoder based on mixed-integer programming~\cite{gurobi}.

Given either decoder, we first perform decoding using only the syndromes of the four ancillary logical qubits to infer their logical outcomes.
The syndromes of the output logical qubit are not used at this stage, because the factory postselection should be done without measuring the output logical qubit, so that the output can continue to be used in subsequent logical operations.
We perform factory postselection on the distillation logical outcome being $0011$, which is the desired outcome for our distillation circuit.
We then optionally perform further stabilizer postselection, which can further boost the fidelity of the output magic state by flagging bad executions of the distillation circuit.
After postselection, we decode with the full syndrome of all five logical qubits to infer the logical outcome of the output magic state (with no further postselection), which we use to compute the output fidelity.
We note that agreement between the results of the two rounds of decoding (four vs. five logical qubits) could be further utilized to herald logical errors in the execution of the full circuit.

The MLD decoder is only tractable for $d\!=\!3$, where the number of syndrome combinations is limited.
To construct the MLD decoder, we sample $10^9$ measurement samples for our full 35-qubit circuit, under the noise model described above.
Our lookup table $T$ will have $2^{15}$ keys corresponding to all possible syndromes, each key storing $2^5$ entries corresponding to the number of occurrences of each logical observable pattern among our samples with the given syndrome.
Sampling can be done efficiently in Stim, by replacing the input magic states with stabilizer states (see our noise learning method described above). 
With each sample, we store the 15-bit syndrome information and the 5-bit logical error string into $T$.
After all samples are collected, each syndrome $s$ will have a most likely logical error $\ell_s$, which will be our decoder output for $s$.
To perform postselection, we can sort the stabilizer patterns based on the logical fidelity of the output they lead to, and perform sliding scale postselection based on this.

The MLD decoding method described in the previous paragraph only works for small code distances, because the space complexity for the table is exponential. Therefore, it is not realistic to use it for decoding at $d\!=\!5$. We therefore use a most-likely error (MLE) decoder, adapted from Ref.~\cite{cain2024correlated,landahl2011fault}, to decode the logical measurement results and evaluate their confidence for post-selection.

We construct an MLE decoder based on mixed-integer programming (MIP) formalized as follows. We denote all stabilizers as $\Sigma=\{ \sigma_{1},\dots, \sigma_{k} \}$, and all logical observable as $O=\{ O_{1},\dots,O_{l} \}$. We enumerate all possible elementary Pauli errors $\mathcal{E} =\{ \epsilon_{1}, \dots, \epsilon_{m} \}$ in the injection and distillation circuits, and each error $\epsilon_{j}$ can flip a subset of stabilizers $\Sigma_{j} \subset \Sigma$ as well as a subset of logical observables $\Omega_{j} \subset O$ with probability $p_{j}$. If we define \begin{equation}
    \partial_{i,j} = 1\text{ if } \sigma_{i}\in \Sigma_{j},  L_{i,j} = 1\text{ if } O_{i}\in \Omega_{j},
\end{equation}
then given an error configuration $\vec{e}\in \mathbb{F}_{2}^m$, the resulting stabilizer and observable configuration will be $\partial \vec{e}$ and $L \vec{e}$, respectively.
The input of the MLE decoder is a stabilizer configuration $\vec{s}=\begin{pmatrix} s_{1} \\ \vdots \\ s_{k} \end{pmatrix} \in \mathbb{F}_{2}^k$, and it will return the most-likely error configuration that results in the same stabilizer configuration. More precisely, the most-likely error is defined \begin{equation}
    \vec{e} = \mathrm{argmax}\ P(\vec{e}),\text{ s.t. } \partial \vec{e} = \vec{s},
\end{equation}
where $e_{j},s_{j}$ are binary variables. Equivalently, the most-likely error can be determined by the following mixed-integer programming problem by regarding all variables as integers and  introducing new slack variables $\lambda_{j}$: \begin{equation}
    \vec{e} = \mathrm{argmax}\sum_{j} \log\frac{1-p_{j}}{p_{j}} e_{j},\text{ s.t. } \sum\partial_{i,j} e_{j} = s_{j} + 2\lambda_{j},
\end{equation}
where $e_{j}, s_{j}, \lambda_{j}$ are integers.

To post-select a shot based on the stabilizer configuration, we analyze the logical gap~\cite{bombin2022fault,meister2024efficient,smith2024mitigating,gidney2023yoked}, which characterizes the confidence in the chosen correction. We seek to characterize the confidence by analyzing the likelihood of this error compared to those resulting in other logical corrections. We define the second-most-likely error (SMLE) to be \begin{equation}
    \vec{f} = \mathrm{argmax}\ P(\vec{e}),\text{ s.t. } \partial \vec{f} = \vec{\sigma}\text{ and } L\vec{f}\neq L \vec{e},
\end{equation}
then the logical gap of a given stabilizer configuration is defined as \begin{equation}
    g = \log{\frac{P(\vec{e})}{P(\vec{f})}}.
\end{equation}
Intuitively, the logical gap provides a confidence measure for decoding---the gap approximates the likelihood difference between the most likely logical outcome and the second most likely logical outcome.

In the case of 5-to-1 distillation, there are four logical qubits that are measured, and we will use the stabilizer information from those four to post-select the shots. We enumerate all $2^4$ logical representatives over these four measured qubits and add the corresponding logical observable as a new constraint into the MIP solver to obtain the MLE and the SMLE.
To determine whether we accept a shot, we compute the logical gap based on the detector information on the measured four logical qubits, and see if it is greater than a logical gap threshold we set ahead of time.

We observe that for $d=3$, the logical error performance for the MLD and MLE are comparable (Fig.~\ref{fig:d3_distillation}b).
This suggests that the additional entropic contribution from considering all error cosets is smaller than that coming from analyzing the most likely error itself.

\methodssubsection{Physical error rate to achieve distillation gain without stabilizer postselection}
We now perform numerical simulations of our $d=3$ distillation process at a variety of different physical error rates, and evaluate the injected and distilled magic state fidelities in the absence of stabilizer postselection.
This provides an estimation of how much the physical error rate should be improved to see distillation gain without extra postselection penalties, and future work can extend this to a comparison between different code distances.

The results are shown in Extended Data Fig.~\ref{fig:decoder_cuts}a.
As we globally rescale the physical error rate, both the injected and distilled fidelity improve, with the distilled fidelity improving faster due to its quadratic scaling.
To match the experimentally-observed fidelities (crosses), the physical error rates are rescaled by 1.25$\times$ compared to the error model used for decoding.
We find that an approximately two-fold improvement in physical error rate suffices to achieve distillation gain without stabilizer postselection.